\documentclass[twocolumn,showpacs,prl,preprintnumbers,amsmath,amssymb]{revtex4}

\usepackage{graphicx}
\usepackage{dcolumn}
\usepackage{bm}

\begin{document}
\title{Magnetic excitations in the spinel compound Li$_x$[Mn$_{1.96}$Li$_{0.04}$]O$_4$ ($x$= 0.2, 0.6, 0.8, 1.0):
how a classical system can mimic quantum critical scaling}

\author{Thomas Heitmann$^1$, Alexander Schmets$^{2,3}$, John Gaddy$^{1,2}$, Jagat Lamsal$^{1,2}$,
Marcus Petrovic$^2$,
Wouter Montfrooij$^{1,2}$, and Thomas Vojta${^4}$}
\affiliation{$^1$The Missouri Research
Reactor, University of Missouri, Columbia, MO 65211\\
$^2$Department of Physics and Astronomy, University of Missouri, Columbia, MO 65211\\
$^3$Reactor Institute Delft, Technical University of Delft, 2629 JB
Delft, the Netherlands\\
$^4$Department of Physics, Missouri
University of Science and Technology, Rolla, MO 65409}
\begin{abstract}
We present neutron scattering results on the magnetic excitations in
the spinel compounds Li$_x$[Mn$_{1.96}$Li$_{0.04}$]O$_4$ ($x$= 0.2,
0.6, 0.8, 1.0). We show that the dominant excitations below $T \sim$
70 K are determined by clusters of Mn$^{4+}$ ions, and that these
excitations mimic the $E/T$-scaling found in quantum critical
systems that also harbor magnetic clusters, such as
CeRu$_{0.5}$Fe$_{1.5}$Ge$_2$. We argue that our results for this
classical spinel compound show that the unusual response at low
temperatures as observed in quantum critical systems is (at least)
partially the result of the fragmentation of the magnetic lattice
into smaller units. This fragmentation in quantum critical systems
is the direct and unavoidable result of intrinsic disorder.
\end{abstract}
 \maketitle
\section{introduction}
Metals that harbor magnetic ions can exhibit long range magnetic
order \cite{longrange,longrange1}. The interaction mechanism
responsible for this ordering is the Ruderman, Kittel, Kasuya, Yosida (RKKY) interaction \cite{rkky1,rkky2,rkky3,rkky4}; this mechanism describes how
conduction electrons get polarized by atomic magnetic moments, and
how they in turn can line up neighboring moments. The strength of
the interaction depends both on the degree of overlap between the
localized electronic states and the extended conduction electron
states, as well as on the distance between the neighboring moments.
Since the interaction is mediated by the conduction electrons, the
ensuing pattern of magnetic ordering reflects the shape of the
Fermi-surface  rather than the crystal lattice,
resulting in incommensurate ordering.\\

By compressing the crystal lattice, the degree of hybridization
between the localized and extended states can be tweaked
\cite{doniach}, resulting in a change to the magnetic ordering
tendencies. In fact, if the hybridization becomes very strong, the
conduction electron will become virtually localized and will end up
shielding the magnetic moments. This Kondo-shielding mechanism can
thus destroy magnetic order \cite{doniach}, resulting in a metal
where the conduction electrons and the magnetic moments have
combined to form heavy quasi-particles \cite{sachdev,stewart}. This
heavy-fermion state is reflected in the macroscopic properties
\cite{stewart} such as
the specific heat of the metal.\\

\begin{figure}[t]
\includegraphics*[viewport=0 0 430 360,width=80mm,clip]{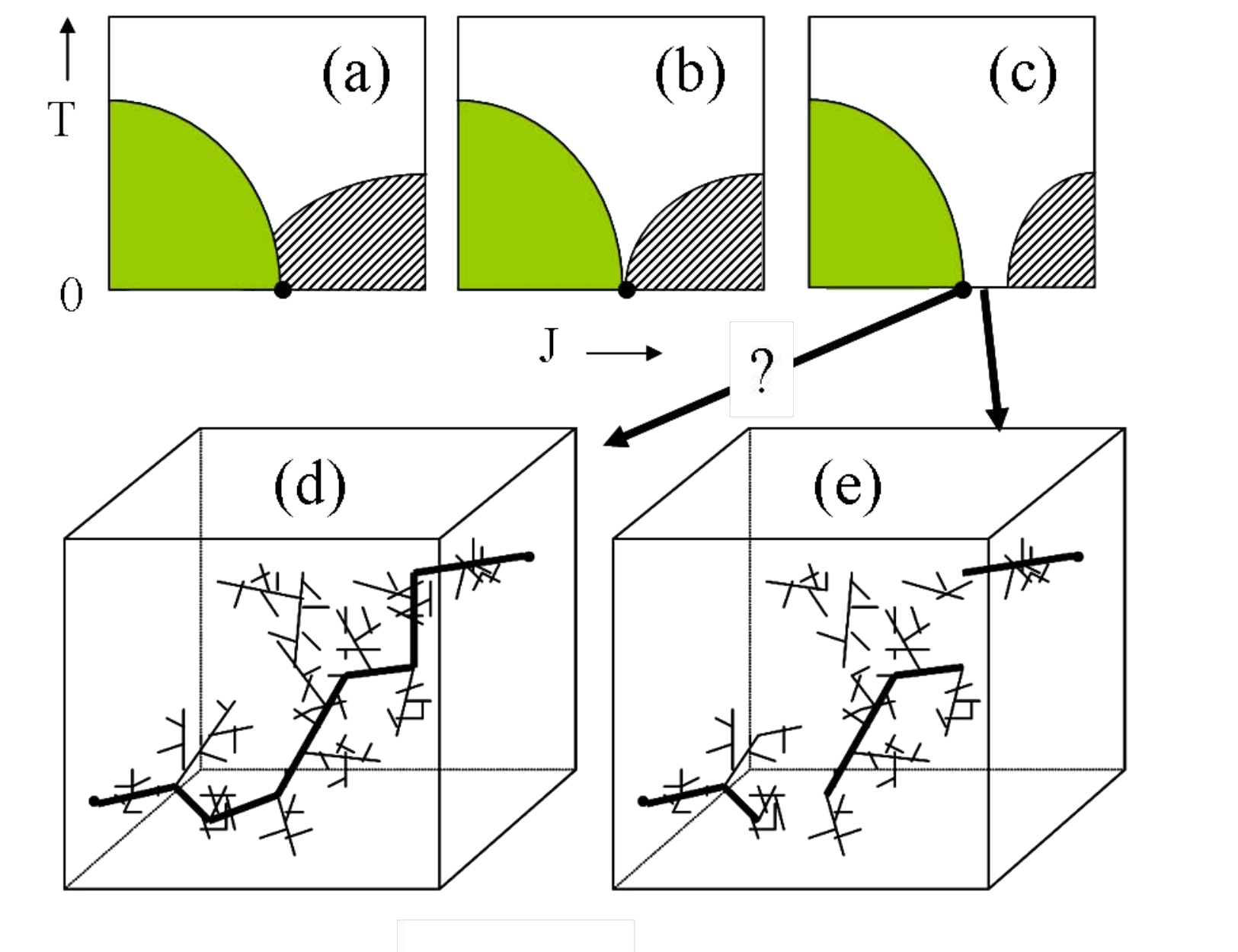}
\caption{(color online) Figure reproduced from Ref.
\cite{montfrooij1}. The phase diagram near the QCP (black dot) as a
function of the coupling strength $J$ between magnetic moments and
conduction electrons. This coupling strength can be tweaked by
applying hydrostatic or chemical pressure. The striped areas
indicate that the moments are fully shielded, the solid areas are
the region where long-range order is present. (a) The Hertz-Millis
scenario \cite{hertz,millis} where ordering is established through a
spin-density wave instability of the Fermi surface. (b) The local
moment scenario \cite{schroeder,si} where the QCP is the point where
moments can first survive, with decreasing $J$, all the way down to
0 K. (c) Local variations in interatomic distances has resulted in a
spread in shielding temperatures yielding a region in between the
areas where the moments order and where they are fully shielded
\cite{montfrooij1}. In this region the surviving (down to 0 K)
moments are located in clusters that are not interconnected, as
shown in part (e). In this latter scenario, the QCP would correspond
to the presence of a lattice spanning cluster, as depicted in part
(d).} \label{phasediagram}
\end{figure}

By applying just the right amount of lattice compression (or
expansion), the metal can be tweaked to such an extent that the
system will be on the verge of magnetic ordering at T = 0 K
\cite{doniach,sachdev,stewart}, while the magnetic moments are
completely \cite{hertz, millis} (see Fig. \ref{phasediagram}a), or
almost completely \cite{schroeder} (Fig. \ref{phasediagram}b)
shielded. The metal is said to be at the quantum critical point
(QCP), and its macroscopic properties (specific heat, resistivity
and susceptibility) are no longer described by Fermi-liquid theory.
This non-Fermi liquid (nFl) state of matter \cite{sachdev,stewart}
can also be accompanied by dynamic scaling laws \cite{aronson}. For
instance, when the system is being probed on a microscopic scale such as
is being done in neutron scattering experiments, its response to
a perturbation from equilibrium by an amount of energy
$E$ while at a temperature $T$,
depends only on the ratio $E/T$ \cite{aronson,schroeder,meigan,montfrooij}.\\

The emergence of $E/T$-scaling was particularly surprising since it
occurred not only in systems that were (most likely) just below
\cite{si} the upper critical dimension \cite{sachdev}
(CeCu$_{5.9}$Au$_{0.1}$ \cite{schroeder}), but also in systems that
were manifestly above the upper critical dimension
(UCu$_{5-x}$Pd$_{x}$ ($x$= 1.0, 1.5) \cite{aronson,meigan} and
Ce(Ru$_{0.24}$Fe$_{0.76}$)$_2$Ge$_2$ \cite{montfrooij}). Montfrooij
{\it et al.} suggested \cite{montfrooij1} a way out of this apparent
contradiction based on the disorder present in a system \cite{neto}
when it is driven to the QCP by means of chemical substitution.
Their arguments are straightforward and are summarized in Fig.
\ref{phasediagram}c. When chemical doping takes place, locally the
inter-atomic distances are changed by a very small amount, typically
on the order of 0.05 \AA $^{-1}$. While this is a small number, it
is enough to locally change the overlap between neighboring orbitals
to such an extent that it will introduce a distribution of
temperatures at which the moments become fully shielded. Then, when
the system is cooled down, some moments will be shielded while other
moments will survive (perhaps partially shielded) all the way down
to 0 K. The structure that evolves upon cooling down is a
percolation network that
follows the random chemical doping sites.\\

Note that the emergence of such a percolation network is unavoidable
since in a system that is on the verge of ordering any change, even
very small ones, will make a significant difference. This was shown
in the heavy-fermion compound Ce(Ru$_{0.24}$Fe$_{0.76}$)$_2$Ge$_2$
\cite{montfrooij1}. Starting from the heavy-fermion system
CeFe$_2$Ge$_2$ \cite{cefege}, upon {\it increased} doping $x$ of Ru
on the Fe sites Ce(Ru$_{0.24}$Fe$_{0.76}$)$_2$Ge$_2$ orders
magnetically
 at 0 K once 1:4 Fe ions have been
substituted\cite{fontes,montfrooij}. This iso-valent substitution
takes place on sites that are not nearest neighbors to the moment
carrying Ce-sites. Thus, one could have expected the effects of this
chemical doping to be uniformly spread out over the crystal lattice.
However, instead of observing that all the magnetic moments evolve
with temperature along the lines sketched out in Fig
\ref{phasediagram}a or \ref{phasediagram}b, unambiguous evidence
\cite{montfrooij1} for the emergence of magnetic clusters (Fig.
\ref{phasediagram}c) was observed.\\

The emergence of magnetic clusters in QCP-systems upon cooling
should greatly influence the response of these systems and might
well explain some of the unusual nFl-properties. When a system
changes its morphology from a 3-dimensional system to some sort of a
fractal structure, its response should change accordingly
\cite{staufer}. For example, when a magnetic cluster becomes
isolated from the rest of the system, this cluster might well be
forced to order magnetically since the thermal energy required to
keep this system disordered might no longer be available. After all,
the lowest available magnon energy increases with decreasing cluster
size because the longest wave length of such a disordering
fluctuation cannot exceed the size of the cluster. Thus, cluster
formation would be reflected in the specific heat of the system;
whenever a cluster becomes isolated and is forced to order
magnetically because of finite size effects, the loss of entropy
will show up in the specific heat, bestowing an unusual temperature
dependence upon the specific heat that cannot be described by any
theoretical model \cite{hertz,millis,si} that treats the moment
shielding
as being equivalent on all lattice sites.\\

The observed $E/T$-scaling
\cite{aronson,schroeder,meigan,montfrooij,montfrooij1a} might also
be directly or indirectly linked to the formation of magnetic
clusters. It could be a direct result because the fractalization of
the system could drop the dimensionality of the system to below the
upper critical dimension \cite{sachdev}. Or it might be an indirect
result since the newly formed (and forming upon cooling) clusters
can now act as superspins that have their orientation controlled by
a thermally activated barrier $\sim e^{-\Delta/T}$. And akin to the
emergent excitations of such a cluster 'protectorate' \cite{lee}
observed in the spinel compound ZnCr$_2$O$_4$, the reorientations of
these magnetic clusters could well be the local energy modes most
relevant to the
dynamics near the QCP.\\

In here we investigate whether $E/T$-scaling might indeed be linked
to the formation of magnetic clusters upon cooling. To do so, we
study the classical spinel system
Li$_x$[Mn$_{1.96}$Li$_{0.04}$]O$_4$ ($x$= 0.2, 0.6, 0.8, 1.0) by
means of inelastic neutron scattering experiments. This is an
extension of the work of Lamsal {\it et al.} \cite{lamsal} who
investigated the possibility of $E/T$-scaling in the $x$= 1.0
compound. We show that the dynamics of the magnetic clusters that
are present in these systems \cite{gaddy} below $\sim$ 70 K do
indeed mimic $E/T$-scaling behavior for all concentrations $x$
ranging from geometrically frustrated short-range magnetism
($x$=1.0), to long-range ordering ($x$= 0.2). Thus, we show that the
behavior previously associated with quantum critical behavior
ascribed to the competition between ordering and shielding
tendencies, is instead most likely to be strongly influenced by the
emergence of magnetic clusters in these QCP-systems.\\

\section{Experiments and results}
Stoichiometric LiMn$_2$O$_4$ undergoes a charge-ordering (CO)
transition at $\sim$ 300 K \cite{battery} that is accompanied by a
transition from a cubic to an orthorhombic structure. The CO
structure of the Mn$^{3+}$ and Mn$^{4+}$ B-site ions in the
insulating phase has been resolved by Rodriguez-Carvajal {\it et
al.} \cite{rc}. Upon cooling down further to below 66 K, long range
antiferromagnetic (AF) ordering develops \cite{greedan} despite the
large degree of geometric frustration inherently present in a
lattice that has spins located at the vertices of corner-sharing
tetrahedra. The low temperature magnetic
structure has not been resolved yet.\\

Upon doping ($y$) with Li on the Mn-sites,
Li$_x$[Mn$_{2-y}$Li$_{y}$]O$_4$ becomes a material that has
applications as a component of a light weight  battery
\cite{battery}. When a small amount $y$ of Li is substituted on the
Mn-sites, the material retains its capacity for removal of Li from
the A-sites without affecting the overall spinel structure [hence
its use in lithium batteries], but the $\sim$300 K structural phase
transition no longer takes place \cite{battery1}, even though the
CO-transition is unaffected. This suppression of the structural
transition greatly enhances the lifetime of the battery material
during charging/discharging cycles \cite{battery}. Neutron
scattering studies \cite{battery1} and muon investigations
\cite{kaiser,sugiyami} on the doped material have shown that long
range magnetic order is destroyed for $y > 0.02$ and $x > 0.3$, and
the material appears to enter a spin glass phase around 15-25 K
(with the transition temperature depending on the exact amount of Li
substitution \cite{sugiyami}). For the compound with $y$ = 0.04,
long-range magnetic order is re-established below $\sim$ 35 K when the
A-sites are heavily depleted ($x$ = 0.2). Note that the long-range
ordering pattern found for $x$ = 0.2 is not the same as the
long-range ordering pattern observed \cite{greedan} in
stoichiometric LiMn$_2$O$_4$ below 66 K, instead
it is described by the structure established \cite{greedan1} in $\lambda$-Mn$_2$O$_4$.\\

\begin{figure}[t]
\includegraphics*[viewport=60 200 570 440,width=82mm,clip]{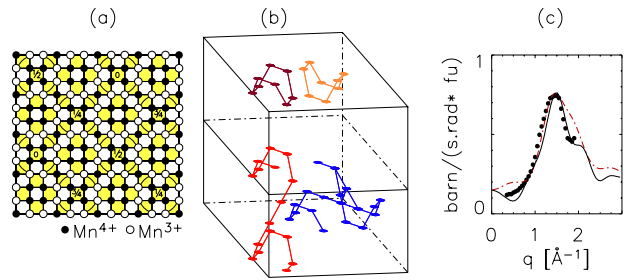}
\caption{(color online) (a) The unit cell for stoichiometric
LiMn$_{2}$O$_4$ projected along the c-axis. The Mn$^{4+}$ ions (open
circles) form isolated 8-fold rings \cite{rc}, while the Mn$^{3+}$
ions (filled circles) form columns along the c-direction. The
numbers within the rings denote their z-position in the unit cell
(b) Two unit cells stacked along the c-axis. Two isolated ring
clusters are shown at the top of the figure. Upon Li substitution on
the Mn-sites (or Mn removal), some Mn$^{3+}$-ions will become
Mn$^{4+}$-ions, leading to modified rings \cite{gaddy} and linked
rings, two possibilities of which are shown in the bottom half of
the figure. (c) The observed scattering in Li[Mn$_{2-x}$Li$_x$]O$_4$
at 30 K \cite{schimmel} (circles). The 2-ring clusters ( linked
along the c-direction; solid line, bottom left cluster in Fig.
\ref{clusters}b), or linked along the ab-direction; dashed line,
bottom right cluster in Fig. \ref{clusters}b) give a satisfactory
description \cite{gaddy} of the neutron data.} \label{clusters}
\end{figure}

\begin{figure}[b]
\includegraphics*[viewport=70 50 500 560,width=70mm,clip]{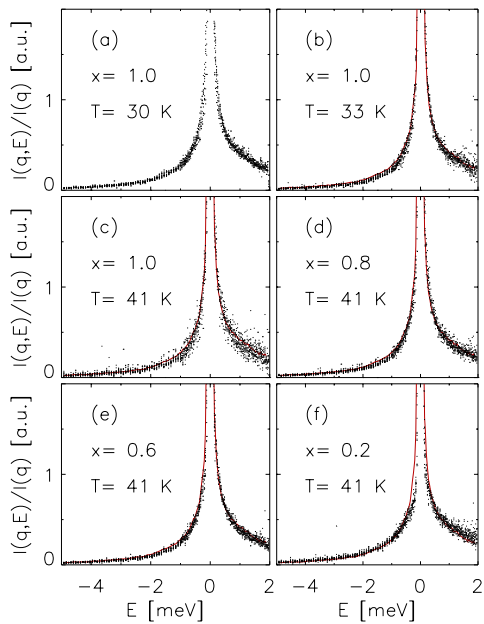}
\caption{(color online) (a) The energy dependence of the scattered
intensity $I(q,E)$ for 13 q-values 0.5 $< q <$ 1.8 \AA$^{-1}$ at T=
30 K
 for the pilot experiment \cite{schimmel}. The energy resolution is given by the sharp
central line due to incoherent nuclear scattering. The data have
been normalized to the magnetic form factor $M(q)$ of the clusters,
which has been determined from $M(q)=I(q,E=0.6)$ \cite{explain}. All
data lie on a single curve, implying that the cluster dynamics are
independent of $q$. (b)-(f) Same as part (a), but now for the new
ILL data taken at $x$= 1.0, 0.8, 0.6 and 0.2, respectively at T= 33
K [part (b)] and at T= 41 K (parts (c) through (f)]. The solid lines
in part (b)-(f) are the smoothed data from part (a) to show the
consistency between the data sets at T $\sim$ 30 K and to provide a
reference for T= 41 K. The scatter of the data points in part (c) is
because the particular experimental run was shorter than the ones
shown in the other parts of the figure.} \label{scaling1}
\end{figure}

Recently, Gaddy {\it et al.} \cite{gaddy} have shown that the
short-range order observed in Li$_x$[Mn$_{1.96}$Li$_{0.04}$]O$_4$
($x$ =1.0) corresponds to the presence of magnetic clusters of
Mn$^{4+}$-ions. Their observations were based on the observations by
Schimmel {\it et al.} \cite{schimmel} that the dynamics associated
with Mn$^{4+}$-ions are much slower than those of the
Mn$^{3+}$-ions; below $\sim$ 25 K \cite{kaiser,schimmel} the spin
dynamics of the Mn$^{4+}$-ions are so slow that the spins appear to
be frozen. The Mn$^{4+}$-ions are the ones that couple
antiferromagnetically to each other, and line up with their
immediate neighbors below $\sim$ 70 K to form the magnetic clusters.
These clusters weakly interact with each other, while the overall
spin of the cluster can reorient itself. The Mn$^{3+}$-ions do not
appear to partake in any cluster formation. The backbone of these
magnetic clusters (see Fig. \ref{clusters}a) are formed by 8-fold
rings of AF-coupled Mn$^{4+}$-ions \cite{rc} corresponding to the
CO-pattern of the stoichiometric compound. We summarize these
observations in Fig. \ref{clusters}. The Li substitution on the
Mn-sites changes some of the Mn$^{3+}$-ions to Mn$^{4+}$ ions,
producing links between the 8-fold Mn$^{4+}$-clusters \cite{gaddy}
(see Fig. \ref{clusters}b). The dynamics of the clusters do not
display any particular length scale dependence (see Fig.
\ref{scaling1}a) as evidenced by plotting the scattering data for
all momentum transfers $\hbar q$, scaled by magnetic form factor of
the Mn$^{4+}$ clusters. The fact that all curves coincide implies
that the dynamics are independent of probing wave length $\lambda$
in the range 0.5 \AA$^{-1} < q=2\pi/\lambda < 1.8 $ \AA$^{-1}$.
Since this is the anticipated behavior for very weakly interacting
superspins, Gaddy {\it et al.} took this as further evidence that
the low temperature ($T <$ 70 K) dynamics in the $x$ =1.0 system are
dominated by the dynamics of the clusters,
similar to the case of ZnCr$_2$O$_4$ \cite{lee}.\\

\begin{figure}[t]
\includegraphics*[viewport=60 115 470 300,width=82mm,clip]{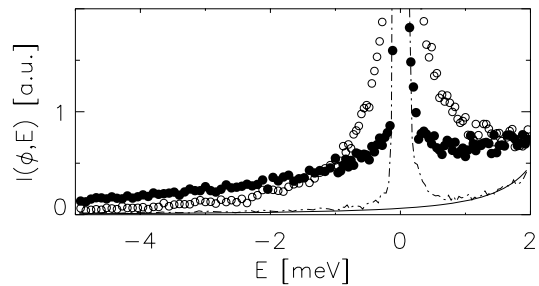}
\caption{Details of representative data sets for the new ILL
experiment showing the total scattering for $x$= 1.0 between 60 $<
\phi <$ 80 degrees at 150 K (filled circles), at 41 K (open circles)
the empty container scattering (dashed-dotted curve) and the time
independent background (solid line) as obtained from an empty
spectrometer run. As can be seen, the container scattering is
negligible for $|E| >$ 0.5, and the data correction only requires
subtraction of the time independent background for $|E| >$ 0.5, the
range we used to test dynamic scaling laws.}\label{ILLraw}
\end{figure}

We have performed new neutron scattering experiments on
Li$_x$[Mn$_{1.96}$Li$_{0.04}$]O$_4$ (with $x$ =1.0, 0.8, 0.6 and
0.2) using the IN6 time-of-flight spectrometer at the Institute
Laue-Langevin (ILL) and using the TRIAX triple-axis spectrometer at
the Missouri Research Reactor (MURR). The setup for the ILL
experiments was identical to the pilot experiments \cite{schimmel} on
Li[Mn$_{1.96}$Li$_{0.04}$]O$_4$; the spectrometer was operated using
a fixed neutron incident energy of 3.12 meV. The amount of sample
for each $x$ was about 1.5 g, and each sample was measured for about
12 hours for a series of temperatures between 2 K and 320 K. Empty
container runs at 50 K showed that most of the container background
was restricted to the elastic channels (see Fig. \ref{ILLraw}), and
that the time independent background was the same as in the pilot
experiments. The quasi-elastic channels, which we use in our search
for $E/T$-scaling, were corrected for this time independent
background. Since the amount of sample was very small compared to
the pilot experiments where 25 g was used, we did not need to
correct the data for any attenuation effects (which were already
small for the 25 g pilot experiment).\\
\begin{figure}[b]
\includegraphics*[viewport=40 50 600 380,width=90mm,clip]{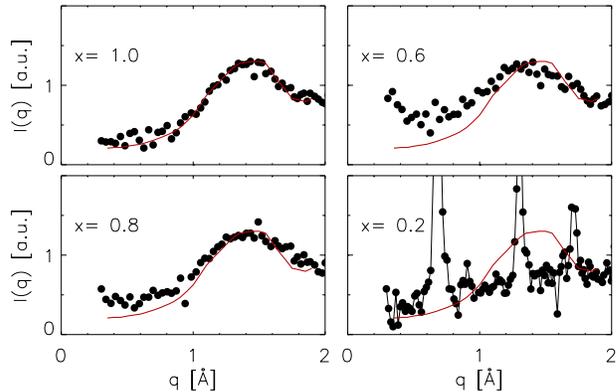}
\caption{(color online) The scattering (circles) associated with the
magnetic clusters for four compositions given in the figure. This
scattering, determined from the total magnetic scattering at T= 24 K
in the energy window $|E| < 0.15$ meV represents the slowest- or
even frozen in- dynamics of the clusters, yielding the magnetic form
factor of the clusters. The almost fully lithium depleted sample
($x$= 0.2) displays long-range order (sharp Bragg peaks
\cite{greedan1}), as well as a substantial amount of short-range
order. The data have been normalized to sample weight, but an
absolute normalization was not possible. For reference, the
absolutely normalized data at $x$ = 1.0 (scaled to give the best
agreement with the new ILL data) obtained from the pilot experiment
\cite{schimmel} are shown by the solid lines in each
figure.}\label{frozen}
\end{figure}

In order to determine the frozen-in component of the magnetic
scattering -which yields the magnetic form factor of the clusters-
for temperatures below 100 K, we subtracted the total scattering
$I$($\phi$,$|E| < 0.15$) at T=100 K at a given scattering angle
$\phi$ from the intensity measured in the same energy window at $T
$= 24 K. This temperature was chosen to be reasonably close to the
glass transition, yet high enough so that the slowing dynamics of
the Mn${3+}$ ions would not interfere with the determination of the
magnetic form factor of the Mn$^{4+}$-clusters. No further data
correction was required for the ILL data sets, and comparison
between the new results for $x$ =1 .0 and the pilot experiments
\cite{schimmel} at $x$= 1.0 gave identical outcomes (see Figs.
\ref{scaling1} and \ref{frozen}). We show the results for the
frozen-in components in Fig. \ref{frozen}. In this figure one can
see the evolution of the Mn$^{4+}$ clusters upon addition of more
Mn$^{4+}$ ions to the lattice. The clusters gradually change from
the ones determined for the fully lithiated sample \cite{gaddy} ($x$
= 1.0) to the long-range order visible in the almost fully
delithiated sample ($x$ = 0.2). The positions of the magnetic Bragg
peaks seen in the $x$ = 0.2 sample (Fig. \ref{frozen}) correspond
exactly to the magnetic structure resolved for $\lambda$-Mn$_2$O$_4$
by Greedan {\it et al.} \cite{greedan1}. We have not attempted to
resolve the short-range magnetic structure for the intermediate
concentrations $x$= 0.6 and $x$ = 0.8; with the addition of more and
more Mn$^{4+}$-ions, the number of possibilities for the various
cluster shapes becomes too large to try to fit to any particular
cluster shape. Similar to the fully lithiated case \cite{lamsal},
the dynamics do not show any length scale dependence (see Fig.
\ref{scaling1}). Thus, we can safely conclude that clusters are
still present. In fact, as is clear from Fig. \ref{frozen}, even the
$x$ = 0.2 sample shows a large amount of short range
order, associated with remnant clusters.\\

\begin{figure}[t]
\includegraphics*[viewport=60 50 500 450,width=90mm,clip]{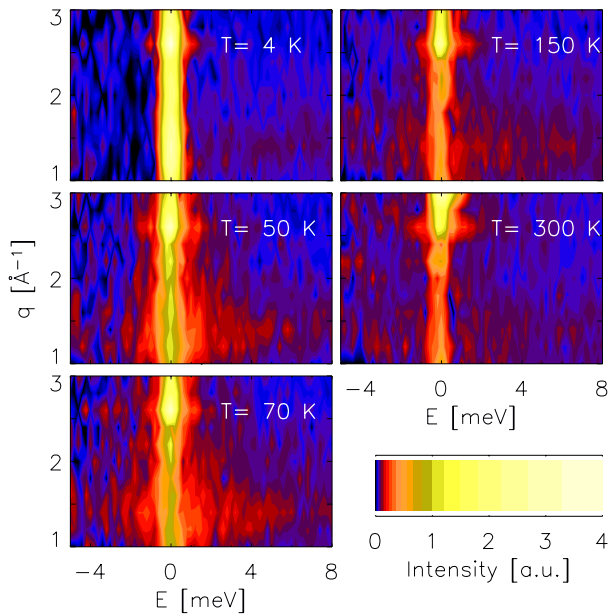}
\caption{(color online) The TRIAX data for $x$= 1.0 after background
correction for temperatures indicated in the figure. The critical
scattering associated with cluster formation around $\sim$ 70 K
\cite{greedan,gaddy} shows up as increased intensity in the
inelastic channels. Note that this critical scattering extends over
a wide range of temperatures, which in turn reflects the frustration
\cite{lee} inherent in systems of this type.}\label{triax}
\end{figure}

The TRIAX experiments for the $x$= 1.0 composition were undertaken
to extend the range of temperatures available for following the
evolution of the magnetic scattering for $T <$ 100 K. For these
experiments, 2 g of powder was placed in a slab sample cell which
was housed inside a closed cycle refrigerator. The spectrometer was
operated at a fixed final energy $E_f$ of 13.7 meV, with sapphire
and PG/Si filters in the incoming beam, and PG filters in the
scattered beam. Empty sample holder runs taken at the same
temperatures as the filled sample holder ones were used to correct
the data for background scattering. The data were also corrected for
the monitor contamination resulting from higher order neutrons
reaching the monitor. These higher order neutrons are being
reflected by the incident beam monochromator and are being counted
by the monitor, but they are being filtered out by the PG-filter in
the scattered beam. This results in the monitor counting more
neutrons than were actually used in the experiment; however, the
correction factor is well known
for TRIAX. We show the fully corrected scattering in Fig. \ref{triax}.\\

\section{discussion and conclusion}

Published data \cite{battery1,schimmel}, as well as our present
neutron scattering experiments show that long-range order does not
emerge for $x$ = 1.0, 0.8 and 0.6 down to 2 K, and that substantial
amounts of short-range order are still left in the $x$ = 0.2
compound that does show long-range order below $\sim$ 33 K. All our
results point towards magnetic clusters of Mn$^{4+}$-ions being
present for all concentrations, and that these clusters dominate the
dynamic response. This makes this family of compounds an ideal
testbed for investigating whether quantum critical scaling laws
\cite{aronson,schroeder,meigan,montfrooij,montfrooij1a} can be
associated with cluster formation in such systems. After all, Li$_x$
[Mn$_{1.96}$Li$_{0.04}$]O$_4$ is a purely classical system (from a
phase transition point of view \cite{sugiyami}) that is disordered
because of geometric frustration and Li/Mn-disorder with clusters of
AF-aligned Mn$^{4+}$-ions [superspins] present below $\sim$ 70 K
\cite{gaddy}. Insulating Li$_x$[Mn$_{1.96}$Li$_{0.04}$]O$_4$ should
not exhibit any type of quantum critical scaling as observed in
metallic quantum critical systems \cite{stewart}, however, next we
argue that its response shows all the hallmarks of $E/T$-scaling;
hallmarks that were thought to be associated exclusively with
quantum criticality. We argue, without using any particular line
shape analysis, that scattering originating from the superspins
effectively mimics quantum scaling behavior, and that this scaling
can be observed even in the compound that displays long-range
ordering
($x$ = 0.2).\\
\begin{figure}[t]
\includegraphics*[viewport=70 60 500 615,width=70mm,clip]{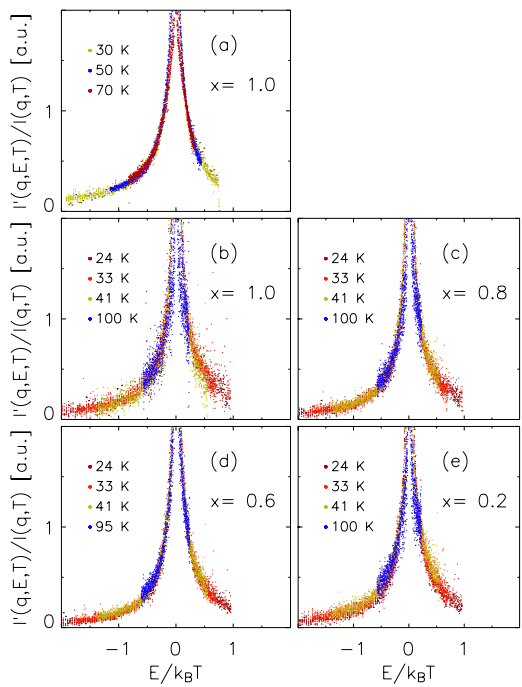}
\caption{(color online) Same as Fig. \ref{scaling1}, but now the
data have been plotted have been multiplied by
$(1-e^{-E/k_BT})/(E/k_BT)$ and plotted versus $E/k_BT$ (circles).
The data in part (a) are the ILL data of the pilot experiment
\cite{schimmel}, parts (b)-(e) correspond to the new ILL
experiments. The Li-concentrations $x$ and temperatures are shown in
the figure. All curves collapse onto each other, giving the
appearance of dynamical scaling. The finite energy resolution
(Fig.\ref{ILLraw}) of the spectrometer causes the apparent
differences between the temperatures for $|E/k_BT| <$
0.1.}\label{scaling}
\end{figure}

Lamsal {\it et al.} \cite{lamsal} have shown that the data taken in
the pilot experiment \cite{schimmel} on the fully lithiated compound
($x$= 1.0) display what looks like $E/T$-scaling for $T$= 30, 50 and
70 K. We extend these observations to all concentrations $x$ in
order to show that the $E/T$-scaling mimicking behavior is present
for all $x$, even when long range order is present at low
temperatures ($x$= 0.2). To demonstrate this, we plot the full
response for $T$ = 24, 33, 41 and 100 K as a function of $E/T$ for
all four concentrations $x$ in Fig. \ref{scaling}. In order to
achieve the scaling between the various temperatures in this figure,
we have first scaled each $q-value$ to the magnetic form factor as
was done in Fig . \ref{scaling1} with the caveat that the form
factor $M(q)$is now determined \cite{explain} from the scattering at
$I(q, E = 0.6 (T/41K))$ [anticipating $E/T$-scaling behavior]. We
also have taken out the thermal population factors. Thus we plot
$I'(q,E,T) = I(q,E,T)(1-e^{-E/k_BT})/((E/k_BT)/I(q, E = 0.6
(T/41K))$. Clearly, all curves coincide to such an extent that
$E/T$-scaling behavior appears to take place. Even at the lowest
temperature ($T$ = 24 K, close to the temperature where the
superspins are starting to freeze out
\cite{kaiser,schimmel,sugiyami}) and at the highest temperature ($T$
= 100 K, above the temperature where the Mn$^{4+}$-ions in the
clusters have fully lined up), the line-shape of the response is
still remarkably similar to the data taken at $T$
= 33 and 41 K.\\
\begin{figure}[b]
\includegraphics*[viewport=20 30 590 315,width=90mm,clip]{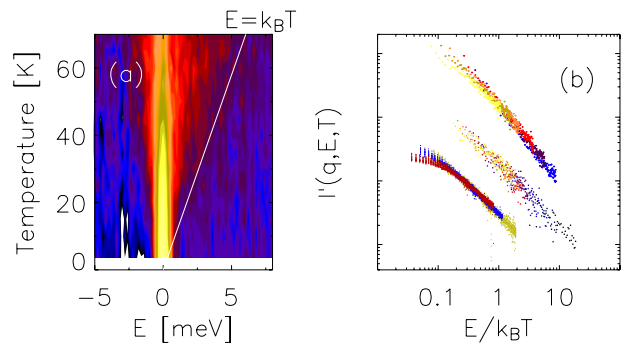}
\caption{(color online) (a) The magnetic scattering at $q = 1.4$
\AA$^{-1}$ for $x$= 1.0 as measured on TRIAX. The scattering behaves
as if $E/T$ is the only relevant variable: the scattering intensity
develops along straight lines $E \sim T$ such as the white line $E
=k_BT$ shown in the figure. (b) When I'(q,E,T) [see text] is plotted
on a log-log scale, the scattering appears to display dynamical
scaling over the full range. The lower curve are the ILL data from
Fig. \ref{scaling}a, the middle curve are all the TRIAX data shown
in part (a), and the upper curve are the data for
Ce(Ru$_{0.24}$Fe$_{0.76}$)$_2$Ge$_2$ at $q$= 0.4 \AA$^{-1}$.
\cite{montfrooij}. The three curves are offset for plotting
clarity.} \label{scaling2}
\end{figure}

In Fig.  \ref{scaling2}a we show the TRIAX data at $q$= 1.4
\AA$^{-1}$ for 5 $< T <$ 70 K. This is the q-value that corresponds
to the strongest scattering for the fully lithiated compound (Fig.
\ref{frozen}). As can be seen in this figure, the scattering appears
to depend only on the ratio of $E/T$, not on $E$ and $T$ as separate
variables. This $E/T$-dependence appears to reach down to the lowest
temperatures, and certainly appears to mimic $E/T$-scaling when
plotted (Fig. \ref{scaling2}b) on a log-log scale
as is usually done \cite{aronson,schroeder,meigan,montfrooij} in QCP-systems.\\

Independent of whether we include the data at $T$ = 100 K and at the
very lowest temperatures or not, it is clear that the $E/T$-scaling
mimicking survives for a wide range of cluster morphologies, even
into the range of compositions where long-range order co-exists with
short-range ordered clusters. This implies that the energy barrier
$\Delta$ that controls the reorientation of the total cluster spin
is likely to be small, a result that was to be anticipated based
upon the findings \cite{lee} in ZnCr$_2$O$_4$. The range of
compositions studied in our spinel system corresponds to a similar
range of compositions investigated for QCP-sytems close to the 0 K
order-disorder transition where isolated magnetic clusters of
various sizes are present \cite{montfrooij1} (Fig.
\ref{phasediagram}e), and would also cover the ordered phase where a
lattice spanning cluster (Fig. \ref{phasediagram}d) is present in
addition to smaller, isolated ones. Therefore it would appear that
in all these cases $E/T$-scaling could be expected based on the
emergence of magnetic clusters alone, independent of any underlying
quantum critical behavior associated with a competition between ordering and shielding.\\

In all, the scattering measured in a classical frustrated magnet
could easily be interpreted as exhibiting $E/T$-scaling such as
observed in CeCu$_{5.9}$Au$_{0.1}$ \cite{schroeder},
UCu$_{5-x}$Pd$_x$ \cite{aronson,meigan} and
Ce(Ru$_{0.24}$Fe$_{0.76}$)$_2$Ge$_2$ \cite{montfrooij}. The dynamics
of this classical system are definitely very similar to what one
observes in chemically doped quantum critical systems. This
similarity is particularly strong in the case of UCu$_{5-x}$Pd$_x$
($x$= 1, 1.5) \cite{aronson,meigan} where the static structure
factor closely follows that of our system, and where the dynamics do
not show any significant q-dependence when plotted as
$S(q,E,T)/S(q,T)$ as we did in Fig. \ref{scaling}. This could very
well imply that magnetic clusters have also formed in
UCu$_{5-x}$Pd$_x$, making it much harder to distinguish between
classical physics and a quantum critical response.\\

We can extend the above in detail to
Ce(Ru$_{0.24}$Fe$_{0.76}$)$_2$Ge$_2$. This latter system displays a
distinct incommensurate ordering wave vector \cite{montfrooij1}, and
might therefore show behavior different from UCu$_{5-x}$Pd$_x$ where
a cluster-like structure factor is evident. However, when we plot
(Fig. \ref{scaling2}b) the response of this system for $q$= 0.4
\AA$^{-1}$ (corresponding to the ordering wave vector
\cite{montfrooij,montfrooij1}) for a range of temperatures [$T$
=1.9, 2.9, 4.4, 7.5, 10.4, 15.4 K] in the same way as was done in
the literature \cite{replot}, then we find that this quantum
critical scaling curve closely resembles that of our classical
system.\\

In conclusion, we have shown that the dynamical response of the
magnetic clusters in Li$_x$[Mn$_{1.96}$Li$_{0.04}$]O$_4$ mimics
$E/T$-scaling behavior. Similar magnetic clusters are expected to be
present \cite{montfrooij1} in quantum critical systems that have
been driven to the QCP by means of (substantial) chemical doping,
and therefore, the $E/T$-scaling observed in these latter systems is
expected to be (strongly) influenced by the emergence of magnetic
clusters. Our findings apply both to quantum critical systems that
seem to be dominated by local physics (UCu$_{5-x}$Pd$_x$) as well as
to systems that show incommensurate long-range order
(Ce(Ru$_{0.24}$Fe$_{0.76}$)$_2$Ge$_2$). We suggest that these
systems should no longer be assumed, a priori, to behave as a
collection of magnetic moments that all undergo shielding at the
same temperature \cite{doniach}. Our results provide a way to
reconcile both possible scenarios for a quantum phase transition (as
depicted in Figs. \ref{phasediagram}a and \ref{phasediagram}b) with
the measured data, by modifying these phasediagrams to include the
effects of chemical disorder that locally leads to lattice expansion
and contraction (Fig. \ref{phasediagram}c). Detailed line shape
analysis of the scaling curves in both classical and quantum
critical systems (Fig. \ref{scaling2}b) has now to be carried out to
be able to determine which part of the response should be attributed
to
geometric effects, and which part to purely quantum mechanical effects.\\

We are currently investigating what influence a distribution of
Kondo-shielding temperatures- leading to cluster formation- would
have on the specific heat of a quantum critical system, and whether
it would be possible that cluster formation might even take place in
QCP-systems that are close to stoichiometric compositions. The
latter proposition is pure speculation at this point; it is based on
the premise that the zero-point motion of the lattice ions would be
sufficient to change interatomic distances by a very small amount,
leading to a distribution of Kondo-shielding temperatures that would
in turn automatically lead to the formation of clusters. The
morphology of these clusters would change as a function of time.
Whether this process is actually relevant or not to the physics near
the QCP would depend on whether the conduction electrons would move
on a faster time-scale than the
zero-point motion of the ions, or not.\\

This material is based upon work supported by the Department of
Energy under Award Number DE-FG02-07ER46381 and by the University of
Missouri Research Board (Grant No. RB-07-52).


\begin{thebibliography}{10}
\bibitem{explain} We chose the point $E$= 0.6 so that we would be
sufficiently far away in energy from the container scattering (see
Fig. \ref{ILLraw}). The exact location in energy does not matter
since all $q$-values display an identical line shape in $E$.
\bibitem{replot} We have taken the published dynamic structure
factor $S(q,E,T)$, and scaled them according to $I'(q,E,T)=
S(q,E,T)T^{0.5}(1-e^{-E/k_BT})/E$, as was done in Ref.
\cite{montfrooij}.
\end{thebibliography}
\end{document}